# MICROSTRUCTURES DIAGRAM OF MAGNETRON SPUTTERED ALN DEPOSITS : AMORPHOUS AND NANOSTRUCTURED FILMS

V. BRIEN[#,] P. PIGEAT

*Laboratoire de Physique des Milieux Ionisés et Applications, UMR CNRS-UHP 7040,*

*Université Henri Poincaré Nancy 1, Faculté des Sciences et Techniques,*

*Boulevard des Aiguillettes, B.P. 239, F-54506 Vandoeuvre-lès-Nancy Cedex, France*

## ABSTRACT

In order to get homogeneous nanostructured Aluminum Nitride deposits, thin films were grown at room temperature on [001] Si substrates by radio frequency magnetron reactive sputtering. The deposits were analysed by Transmission Electron Microscopy, energy dispersive X-ray spectroscopy and Auger electron spectroscopy. Their microstructure and chemical composition were studied versus the plasma working pressure and the radio frequency power. Systematic analysis of cross views of the films allowed the authors to draw a microstructure/process parameters map. Four microstructural types were distinguished according to the decrease of the deposition rate. One is the well-known columnar microstructure. The second one is made of interrupted columns or fibrous grains. The third one is made of nano-sized particles (size of the particles ranges from 1.7 to 8 nm). The fourth and last microstructure is amorphous. The "deposit morphology – process parameters" correlation is commented on.

[#] For correspondence : please contact V. Brien, Valerie.Brien@lpmi.uhp-nancy.fr,

Tel : + 33 (0) 83 68 49 28, Fax : + 33 (0) 3 83 68 49 33

# 1  Introduction

Thanks to high electromechanical coupling, high thermal conductivity (similar to copper) and high sound velocity, AlN is a material usually integrated in devices, especially in SAW (Surface Acoustic Waves) devices. Optimization of this material for the SAW applications led the community to the synthesis of well crystallized and textured films [1]. Its large and direct energy band gap has also made it interesting for opto-electronic applications (photoluminescence when doped). It is also often used in semiconductor multi-layers as an insulator. Additionally, this dielectric material completes its panel by other properties like : excellent corrosion resistance, a high mechanical strength and a weak thermal expansion coefficient that makes it a very attractive material in terms of further applications in micro/nano-systems or sensor technologies.

The AlN films used so far in microelectronics are [002] oriented columnar crystallized films. The decrease of the size of the different elements constituting the electronic devices makes the study of the evolution of the above quoted physical and chemical properties versus the nanosizing necessary. The question of the role of nanosizing on the properties of AlN led us to find a way to directly obtain nanocrystalline films, exhibiting an homogeneous structure all along its thickness, whose physical properties could then be characterized.

In recent years, the study of the nano-scaling of many different materials have shown these new materials exhibit peculiar and interesting physical properties [2]. More specifically, studies on AlN nanopowders have shown the photoluminescence or absorption spectra are modified by the decrease of the grain sizes [3, 4]. Other works mention that the electromechanical coefficient or piezoelectric property of some other materials ($PbTiO_3$, PZT) have been found to be strongly and positively influenced by the nanosizing of their grains [5,

6]. So, low dimensional structures of AlN in films can undoubtedly be expected to produce new materials exhibiting different and hopefully enhanced physical properties. Some literature works do mention the occurrence of amorphous or nanocrystallized AlN structures [7-12]. On the one hand, most of these works relate that they appear when growing films as adaptation layers reducing lattices mismatches between substrates and crystalline AlN layers and are generally unwanted structures [7-11]. On the other hand, Cibert et al. have deposited by pulsed laser deposition (PLD) accumulation of AlN nanocrystallites. However, the homogeneity of this kind of deposits would still be to be studied by further Transmission Electron Microscopy investigations [12]. Unfortunately, PLD is still not the most adequate technique to produce industrially complex micro- or nano- systems. Other works also mention homogeneous nanocrystalline or amorphous thin films designed for photo, cathodo, electro-luminescence. The occurrence of the amorphous state, however, is linked to the presence of a third element, that is to say to the introduction of a third and big atom (transition metal, lanthanide) [13, 14].

So, as far as we know, no well-controlled homogeneous nanocrystallized AlN films have been made by a deposition technique of any interest to industry.

The aim of this study was to examine whether it may be possible to obtain 2D and or 3D nanostructured Aluminium Nitride films by magnetron sputtering. The goal of this work project is to possess high quality nano-structured samples with excellent surface state, whose grain size can be controlled thanks to a well established synthesis process, in order to discover the influence of the nano-morphology on some of its physical properties. The expected modifications will then be subsequently tested to know how they could be exploited in terms of technological applications. To realize this study, the versatile reactive magnetron sputtering technique usually used in microelectronics manufactories was chosen.

Firstly, this paper will describe the experimental reactive magnetron sputtering set-up that was specially designed and built for the purpose of this study in particular to prepare the deposits under ultra low background pressures. The experimental characterization techniques will be detailed, and the results will consist of the combined influence of the following process parameters : plasma pressure (P) and radio frequency power (W) on the microstructure of the deposited thin films. The suggestions on values of experimental parameters to be chosen to get the adequate films will be given and commented on.

## 2  Experimental procedure

### 2.1  *Synthesis*

The specially designed experimental apparatus consisted of a baking UHV (Ultra High Vacuum) chamber using a classical diffusion-pump. A liquid nitrogen cooling trap was used to absorb water vapour. The obtained background pressure was around $1.10^{-6}$ Pa of $H_2O$ controlled by mass spectroscopy. The chamber was equipped with a radio frequency (13.56 MHz) pulsed magnetron sputtering system specifically designed to work under UHV conditions. In this study, bias voltage was set to 0 Volt. The target disk was made of pure aluminium (purity of 99.99 %), its diameter was 60 mm large, the thickness 3 mm in breadth and its distance from the sample was set to 80 mm. A gas mixture of high purity (99.999 %) Ar and $N_2$ was used for sputtering. The percentage $\alpha$ of $N_2$/Ar in the gas mixture was set to 50 %. During the treatment, a controlled pumping valve and mass-flow controllers were used to keep the total sputtering pressure P constant. The flow rates of argon and nitrogen gas were controlled with MKS Mass-Flo Meters (2 sccm $N_2$, 2 sccm Ar) and the total working pressure was measured using a MKS Baratron gauge. Films were deposited on [001] Si wafers. Substrates were cleaned with basic solvents. The last stage of cleaning was an ultrasonic bath in distilled water. The target was systematically sputter cleaned for 15 min using an Ar plasma to remove the native oxide contamination (cleaning conditions : sputtering pressure P = $5.10^{-1}$ Pa, RF power W = 300 W). Prior to deposition, the chamber set into operation with the chosen plasma conditions for 30 mn so that the reactor holding the plasma would reach its thermal equilibrium.

The substrate was thermally insulated. Its temperature measured during deposition thanks to an adequately placed thermocouple only depends on plasma heating, and was found to stay below 50 °C.

The reactor was equipped with an interferential optical reflectometer used in order to control the thickness of the deposit in real-time. The silicon substrates were fixed onto a 6-fold rotating substrate holder. A diaphragm equipped with a shutter allows us to make six different deposits under different plasma conditions with no return to the atmosphere.

Experimental conditions of samples preparation are displayed in a P, W diagram. This diagram is presented in Fig. 1 where each star stands for a deposited film. P lies in the $5.10^{-1}$ - $20.10^{-1}$ Pa range and W is situated in the 50 - 300 W range. The synthesized films thicknesses were chosen around 200 nm to allow the TEM observations and were found to be all strongly adherent to their substrate.

## 2.2 *Techniques of characterization*

TEM observations were carried out on a Philips CM20 microscope operating at an accelerating voltage of 200 kV. They were all performed matter chips sampled on the films directly grown on the Si substrates by the technique of microclivage. A dark field image (DF), a bright field image (BF) and a selected area diffraction (SAD) pattern of both top views and cross views were systematically recorded for each sample.

Elemental composition of the sputtered AlN films was systematically measured by energy dispersive X-ray spectroscopy (EDXS) or Auger electron spectroscopy (AES). EDSX spectra were recorded by means of an EDAX spectrometer mounted on a CM20 Philips microscope and equipped with an ultra thin window X-Ray detector. The analyses were carried out in nanoprobe mode with a diameter of the probe of 10 nm. Concentration profiles were performed by Auger electron spectroscopy on a Microlab VG MKII using an Ar etching gun VG microprobe EX05. The sensibility factors of the two techniques were determined using AlN and $Al_2O_3$ reference samples.

The thickness and the nature of the films were confirmed *ex situ* by TEM diffraction and by spectrometric ellipsometry. The obtained values of the calculated optical index n of the deposited films are in the 1,8 - 2 range (for the wavelength $\lambda = 632$ nm). As predicted in literature, this index depends on the crystalline morphology of the film.

## 3  Results

The TEM systematic crystallographic characterizations of the samples whose synthesis conditions are displayed in Fig. 1 showed that - all films are very dense and contain no porosity and - the film morphologies may be classified in four types following the deposition rate evolution. These zones are schematized on Fig. 1.

Samples exhibiting classical columnar morphology are in zone 1 (Fig. 2). This zone is a high deposition rate zone. Samples made of nanoparticles are in zone 3 (Fig. 4). Zone 2 contains samples exhibiting an intermediate morphology in between the two previous ones (Fig. 3). In zone 4, growth rates are very low, and samples are amorphous (Fig. 4). On the (P, W) diagram of Fig. 1, iso-deposition rate lines and the rough evolution of the oxygen concentration measured in the samples were drawn. The mentioned iso-deposition rates lines (fig.1) were estimated from the exact deposition rates values calculated for every film. The thickness was measured *ex situ* by TEM. It should be noted the oxygen content increases from

5 atomic % in zone 1 to values above 15 atomic % in zone 3 and 4 following the drop of the growth rate of the films.

Zone 1 is the zone where most of the studies encountered in literature and dealing with microcrystalline AlN film growth are performed [7-11,15-18]. As commonly observed the typical microstructure of this zone is made of dense and juxtaposed columns all along the thickness of the film (Fig. 2(a) and 2(b)). The columns go from the bottom to the top of the films and their width progressively increases with increasing thickness. This can actually be described with the well-known Van der Drift formalism [19]. Distribution of width of columns of one single sample is monomodal and the average width of the columns was found to be as high as P was high and as W was small. The maximum width reached by the columns can be measured on the recorded top view (Fig. 2(c)) and can be compared from one sample to another since the films all have the same thickness. The average size ranges from 6 to 18 nm. It is around 6 nm for the sample located at the very bottom right of the diagram Fig. 1. At the very top on the left of the diagram in the zone 1 nanocolumns widths average is around 18 nm. Fig. 2(d) is the typical electron diffraction pattern recorded in this zone. It can be entirely indexed with the würztite structure thereby proving the film is pure and made of the würztite AlN phase. The absence of some of the diffracted rings on this top view SAD pattern indicates the films are textured. Unlike what is commonly mentioned by many authors who have worked on AlN film growth by the same technique on the same substrate, no initial amorphous adaptation layer between the substrate and the crystalline layer could be detected here [7-11]. The resolution of our images however does not enable to deny the presence of such a layer (High Resolution Transmission Electron Microscopy would have been necessary), but our experiment allows to conclude that if such a layer had grown before the crystallization, its thickness would have had to be inferior to 1.3 nm. Nevertheless, a classical intermediate equiaxed crystallites layer before the columnar crystallisation could be observed. Its thickness could be roughly estimated around 20 nm.

Zone 2 is composed of films whose TEM images show a dense juxtaposition of badly defined fibrous grains or shortened columns (Fig. 3). None of the columns could indeed be going from top to bottom of the films (variation in the angle of illumination of the sample was performed). Grain boundaries, however, are well defined and contain no porosity. This microstructure is the same microstructure as the one described as the transition zone, noted "zone T" by Thornton in the well-known reference diagram displaying the microstructures found in films obtained by evaporation under different conditions [20]. This kind of morphology has also been observed by Hwang et al. [9] during the growth of an AlN film obtained by reactive sputtering in a transition layer. This layer was observed between a supposed "$Al_2O_3$ + amorphous" layer on the substrate and a final AlN columnar layer. The typical zone 2 SAD pattern in Fig. 3(d) demonstrates that zone 2 samples are pure and made of AlN crystallites only.

Films of zone 3 are made of AlN nano-particles dispersed in an amorphous matrix. Indeed, these films exhibit electron diffraction patterns showing a discrete diffuse halo (cf. arrow in Fig. 4(a)) and rings indexed with the AlN würztite structure. Dark field micrographs cross view (Fig. 4(b)) and top view (Fig. 4(c)) made with the first AlN ring, show this morphology is homogeneous on the whole thickness of the films (200 nm). As in zone 1 and 2, no amorphous adaptation initial stage could be detected. Average size of the AlN crystallites is 1.7 nm with a maximal size of 8 nm.

The samples prepared in zone 4 located in the very top left part of the P-W diagram (cf. Fig. 1) are amorphous on the whole thickness of the films (200 nm). These samples exhibit SAD patterns essentially made of halos (cf. Fig. 4(d)) typical of amorphous phases. The corresponding bright field image (Fig. 4(e)) also presents the typical orange skin appearance of amorphous matter. These two micrographs were exposed for short periods. The subsequently recorded dark field image (Fig. 4(f)) required much longer exposure times due to weaker densities of light. This image exhibits few local nanocrystallizations that were not visible on the screen (or on the CCD -Charged Coupled Device- camera) before recording the picture. This implies that nano-crystallization was brought about by electron bombardment during exposure. This phenomenon could be observed even in the event of weak electronic currents.

Usually, as mentioned above, the structures found in zones 3 and 4 (Fig. 4) are encountered in AlN films prepared by PVD magnetron as local adaptation layers. In this study, it has been shown that with adapted process conditions, it is possible to obtain these microstructures on the whole thickness of the films in homogeneous manner for thicknesses up to 200 nm.

### 4. Discussion

This work on AlN deposition by magnetron sputtering confirms that the behaviour law linking the deposition rate α and $\frac{W}{P}$ is still valid on all the domain explored. α roughly decreases as $\frac{\sqrt{W}}{P}$ decreases. The work shows that the structures of the AlN films strongly depend on the process parameters and qualitatively describes this dependence. The size and the morphology of the grains are actually both modified by the process parameters. When $\frac{W}{P}$ decreases the crystallites sizes decrease.

The morphology of the AlN films passes from long columns to shorter ones, to nanograins dispersed in an amorphous matrix, and finally the deposits become totally amorphous. One can define a morphological parameter : the shape grain factor $\gamma = \frac{L}{d}$, ratio of its length L over its width d (γ > 1 is a rod shape, γ = 1 is an equiaxe grain, γ = 0 stands for the amorphous structure by convention). A combined decrease of γ and d thus correspond well to the nano-crystallization trend. The map can be summarised as followed : when $\frac{W}{P}$ decreases γ and d decrease.

The evolution towards amorphization is consistent with what is usually put forward in available physical interpretations of sputtered deposits. For instance, different authors aiming at growing the columnar structure have well described the broadening of X-ray peaks attesting to poorly crystallized films when the pressure P increases [17,23,24]. Indeed, the decrease of $\frac{W}{P}$ leads to a stronger thermalization of the species arriving on the sample [25]. This implies the kinetic energy and the flux of species involved in the growth of the films are minimum. The organisation ability or the mobility of the ad-atoms during the growth decrease. This explains poorer crystallizations when $\frac{W}{P}$ is diminished.

Although nanocrystallized deposits have been obtained, one can not conclude however that nanocrystallization is the only consequence of the drop of energy of species. Indeed, as previously noted, when $\frac{W}{P}$ decreases the deposition rate decreases. These nanocrystallized films are obtained for very low growth rates, and are consequently highly contaminated by oxygen. Fig. 1 shows that the combined decrease of the shape grain factor γ and d follow the increase of the oxygen content. So, the oxygen presence may influence the crystallization of the films.

This hypothesis is actually put forward by different observations in literature. Indeed, von Richthofen et al.'s thorough work presenting the phase diagram Al-O-N shows that high oxygen content (>30 %) samples are amorphous [26]. Other authors using magnetron sputtering technique mention the presence of amorphous layers of different thicknesses

(thicknesses of few nm to ones of more than 100 nm) when growing AlN on Si in a background vacuum of $10^{-4}$ Pa [11]. Subsequent to the growth of these amorphous layers, adaptation layers are observed and finally AlN columnar phase develops. It has been suggested that this final crystallization is governed by the absence of oxygen in the reaction chamber due to its consumption during the amorphous layer growth. In their study, it then seems the presence of the amorphous phase could also here be partially due to the excessive presence of oxygen in the first stages of the deposition. More specifically, the controlled addition of small amounts of oxygen in the feed gas during deposition of AlN/AlNO films by reactive sputtering causes the disappearance of the X-ray peaks recorded on the films [27]. This suggests that oxygen damages their crystallographic order.

In our study, the species building the film are Al, N and O due to residual vacuum. Indeed, in spite of the high quality of the background pressure inside the reactor used for the deposits, one can assess that the residual presence of oxygen of $10^{-6}$ Pa (coming from the residual pressure of $H_2O$) corresponds to a collision rate of oxygen molecules/atoms with the substrate of around $10^{13}$ $cm^{-2}.s^{-1}$. It can be calculated that the subsequent oxidation rate could then reach one third of the deposition rates of AlN in zones 3 and 4 (in these zones d < 0.01 $nm.s^{-1}$). Even in the samples prepared under ultra high vacuum, this calculus offers a plausible explanation for the high oxygen levels measured.

Considering the following thermodynamical data ( $\Delta H_f^°$ ($\alpha$-$Al_2O_3$) = - 1672 $kJ.mol^{-1}$, $\Delta H_f^°$ ($\gamma$-$Al_2O_3$) = - 1655 $kJ.mol^{-1}$, $\Delta H_f^°$ (würztite-AlN) = - 317.68 $kJ.mol^{-1}$) one can see that the arrival of an oxygen atom leads automatically to the formation of an Al-O bond [28].

This oxygen atom can be then trapped inside the films. For very small concentrations, solubility of oxygen in AlN is known to cause point defects or stacking faults known as planar inversion domain boundaries [29-31]. The insertion of oxygen atoms can then prevent the perfect epitaxial growth of AlN. The growth of AlN can thus be perturbed by the presence of oxygen owing to as regular interruptions of epitaxy leading to nanocrystallization or amorphization of the films.

Specific studies need to be carried out to specify the influence of oxygen on the microstructure of the AlN films when they are made with very low growth rates.

## 4  Conclusion

This study shows that it is possible to obtain homogeneously 200 nm thick amorphous or nanocrystallized (either 2D or 3D) AlN thin films by RF magnetron reactive sputtering. To do so, it is necessary to decrease the RF power W, and to increase the sputtering pressure P. Unfortunately, these conditions correspond to very low deposition rates. For these values of deposition rate, the study has shown that even if the films are synthesized under UHV they still contain oxygen.

Today, SAW microdevices using these nanocrystalline AlN thin films are being made for technology assessment. These demonstrators will allow the study of the electromechanical properties of these new crystalline morphologies.

**Acknowledgments**

References

1. M.B. Assouar, M. El Hakiki, O. Elmazria, P. Alnot, C. Tiusan, Diamond and Rel. Mater. 13 (2004) 1111.

2. S.C. Tjong, Haydn Chen, Mater. Sci. and Eng. R 45 (2004) 1.

3. A. Olszyna, J. Siwiec, D. Dwiliński, M. Kamińsak, J. Konwerska-Hrabowsak, A. Sokołowska, Mat. Sci. and Eng., B50 (1997) 170.

4. T. Xie, X.Y. Yuan, Y. Lin, X.X. Xu, G.W. Meng, L.D. Zhang, J. Physics, Condensed Matter, 16 (9) (2004) 1639.

5. E.K. Akdogan A Safari., IEEE Transactions on Ultrasonics and Frequency Control. 47 (2000) 881.

6. C. Miclea, C. Tanasoiu, A. Gheorghiu, C.F. Miclea, V. Tanasoiu, Mechanochemistry and Mechanical Alloying 39 (2004) 5431.

7. Y.-J. Yong Lee, H. S. Kim, J.Y. Lee, Appl. Phys. Lett. 71 (1997) 1489.

8. J. H. Choi, J. Y. Lee, J.H. Kim, Thin Solid Films 384 (2001) 166.

9. B.H. Hwang, C.S. Chen, H.Y. Lu, T.C. Hsu, Mat. Sci. and Eng. A 325 (2002) 380.

10. J.-X. Zhang, Chen Y. Z., Cheng H., Uddin 1., Shu Yuan, K. Pita, T.G.Andersson, Thin Solid films, 471 (2005) 336.

11. W.-J. Liu, S.-J. Wu, C-M Chen, Y.-C. Lai, C.-H. Chuang, J. Cryst. Growth 276 (2005) 525.

12. C. Cibert, F. Tétard, P. Djemia, C. Champeaux, A. Catherinot, D. Tétard, Superlat. and Microstr. 36 (2004) 409.

13. M.L. Caldwell, A.L. Martin, C.M. Spalding, V.I. Dimitrova, P.G. Van Pattern, M.E. Kordesch, H.H. Richardson, J. Vac. Sci. Technol. A 19 (2001) 1894.

14. M.L. Caldwell, P.G. Van Pattern, M.E. Kordesch, H.H. Richardson, MRS Internet J. Nitride Semicond. Res. 6, 13 (2001)

15. I. Ivanov, L. Hultm, K. Järrendahl, P. Mäterbsson, J.-E. Sundgren, B. Hjörvarsson, J.E. Greene, J. Appl. Phys. 78 (1995) 5721.

16. M. Akiyama, T. Harada, C. N. Xu, K. Nonaka, T. Watanabe, Thin Solid films 350 (1999) 85.

17. H. Cheng, Y. Sun, P. Hing, Surf. and Coat. Tech. 166 (2003) 231.

18. O. Elmazria, N.B. Assouar, P. Renard, P. Alnot, Phys. Sta. Sol. (a) 196 (2003) 416.

19. A.S. Gudovskikh, J. Alvarez, J.P. Kleider, V.P. Afanasjev, V.V. Luchinin, A.P. Sazanov, E.I. Terukov, Sensors and actuators A113 (2004) 355.

20. M. A. Auger, L. Vazquez, M. Jergel, O. Sanchez, J.M. Albella, Surf. and Coat. Tech. 180-181 (2004) 140.

21. A. van der Drift, Philips. Res. Rep. 22 (1967) 267.

22. J.A. Thornton, J. Vac. Sci. Technol., 11 (1974) 666.

23. X.H. Xu, H.S. Wu, C.J. Zhang, Z.H. Jin, Thin Solid Films 388 (2001) 62.

24. H. Cheng, Y. Sun, P. Hing, Thin Solid Films 434 (2003) 112.

25. G. Gonzalez, J.A. Freitas, C.E. Rojas, Scripta Mater. 44 (2001) 1883.

26. A. von Richthofen, R. Domnick, Thin Solid films 283 (1996) 37.

27. N.J. Ianno, H.Enshashy, R.O. Dillon, Surf. and Coat. Tech. 155 (2002) 130.

28. D. R. Stull, H. Prophet, JANAF Thermochemical tables, Nat. Stand. Ref. Data Ser. (NSRDS), Nat. Bur. Stand. (USA) 1971.

29. G. A. Slack, L. J. Schowalter, D. Morelli, J.A. Freitas Jr, J. Cryst. growth 246 (2002) 287.

30. A. D. Westwood, R.A. Youngman, M.R. McCartney, A.N. Cormack, M.R. Notis, J. Mater. Res. part I, 10, 5 (1995) 1270.

31. A. D. Westwood, R.A. Youngman, M.R. McCartney, A.N. Cormack, M.R. Notis, J. Mater. Res. part III, 10, 10 (1995) 2573.

List of figures and tables captions :

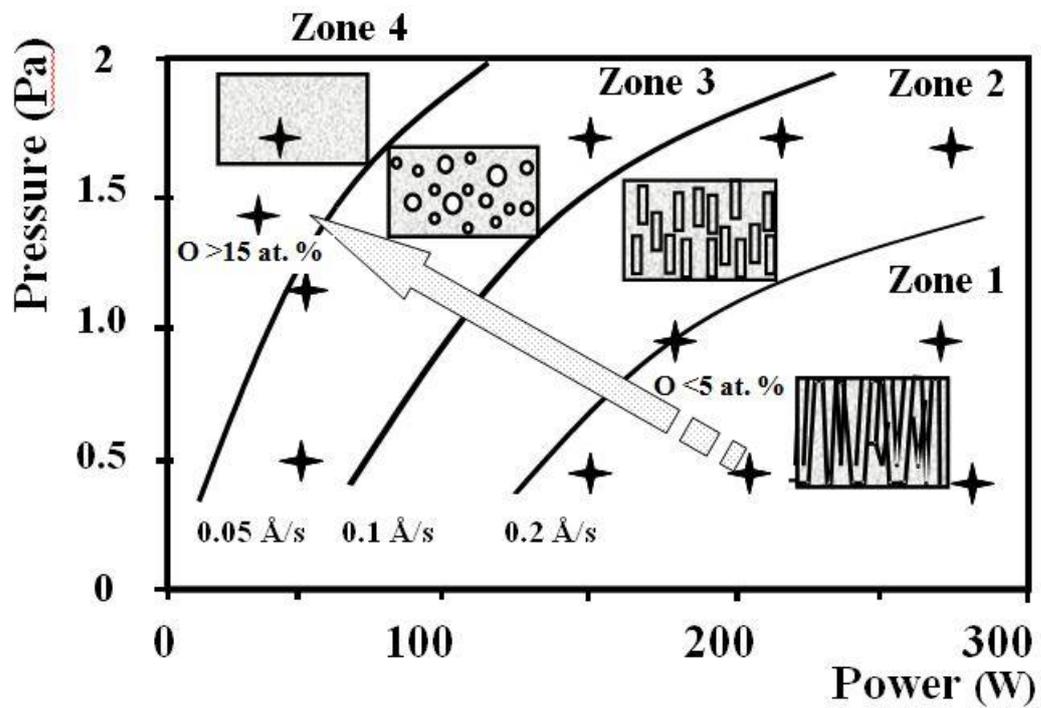

Fig. 1 : Map schematising microstructures morphologies observed by TEM versus the sputtering pressure P and the RF power W. Indicative curves of iso-deposition rate (nm.s$^{-1}$) have been placed. Global evolution of oxygen concentration measured by A.E.S. has been noted (atomic %).

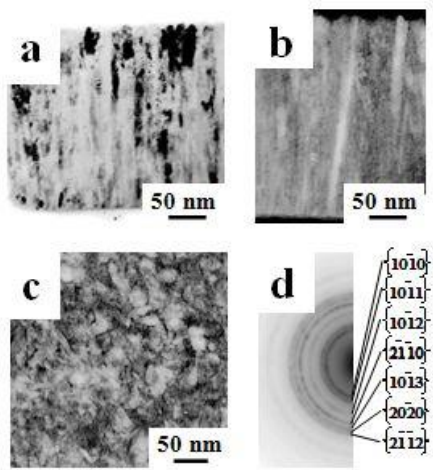

Figure 2

Fig. 2 : Typical TEM micrographs taken on samples in zone 1 of the P-W diagram. Cross views : a/ Bright field image and b/ Dark field image showing the columnar structure of the films. Top views : c/ Bright field micrograph showing the average maximal width of columns, d/ Top view SAD pattern micrograph showing the film is made of the AlN würztite structure.

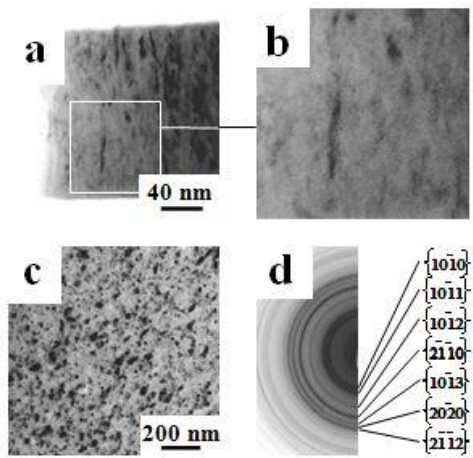

**Figure 3**

Fig. 3 : Typical TEM micrographs taken on samples in zone 2 of the P-W diagram. Cross views : a/ Bright field image, b/ detail of image a. Top views : c/ Bright field image, d/ SAD pattern (AlN würztite indexation).

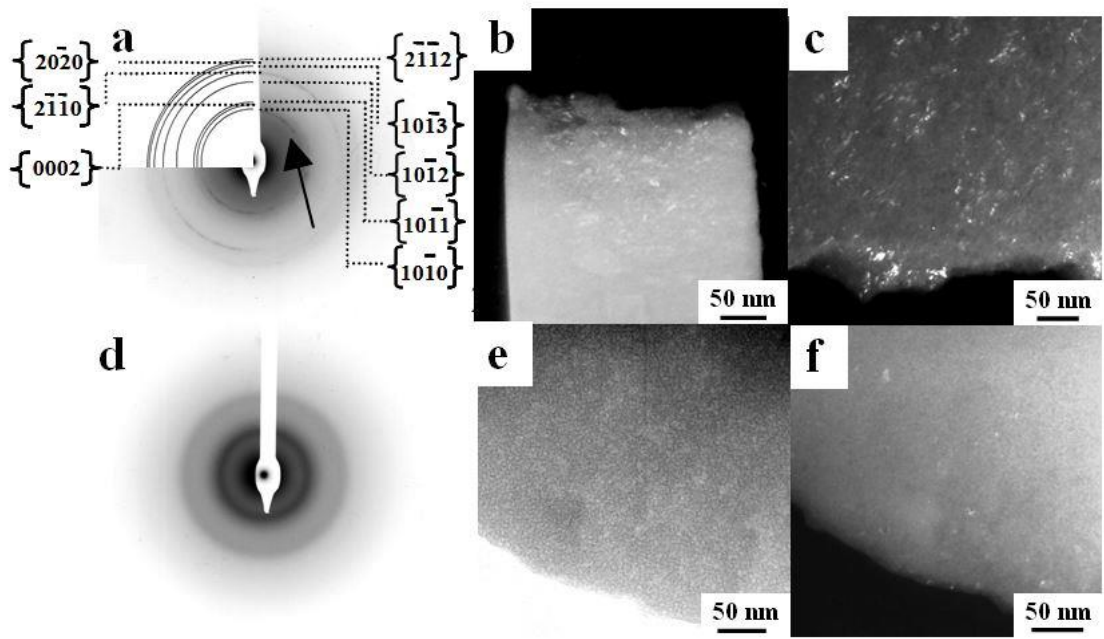

Fig. 4 : TEM micrographs taken on samples of zone 3 and zone 4 of the P-W diagram. Zone 3 : a/ SAD pattern showing the nano-crystallites inside the films are pure and made of AlN würztite, b/ Dark field image of the cross view showing the equiaxed grains structure, c/ Dark field image of the top view. Zone 4 : d/ SAD pattern showing the film is amorphous, e/ Bright field image exhibiting the typical amorphous contrast, f/ Dark field image built with the first amorphous ring.